\documentclass[twocolumn,showpacs,preprintnumbers,amsmath,amssymb,floatfix,elsart]{revtex4}

\usepackage{graphicx}
\usepackage{dcolumn}
\usepackage{bm}

\newcommand{\beq}{\begin{equation}}
\newcommand{\eeq}{\end{equation}}
\newcommand{\bey}{\begin{eqnarray}}
\newcommand{\eey}{\end{eqnarray}}

\begin{document}

\title{Anisotropic strange star with de Sitter spacetime}

\author{Mehedi Kalam}
\email{mehedikalam@yahoo.co.in} \affiliation{Department of
Physics, Aliah University, Sector - V , Salt Lake,  Kolkata
700091, India}

\author{Farook Rahaman}
\email{farook_rahaman@yahoo.com} \affiliation{Department of
Mathematics, Jadavpur University, Kolkata 700032, West Bengal,
India}

\author{Saibal Ray}
\email{saibal@iucaa.ernet.in} \affiliation{Department of Physics,
Government College of Engineering \& Ceramic Technology, Kolkata
700010, West Bengal, India}

\author{Sk. Monowar Hossein}
\email{sami_milu@yahoo.co.uk} \affiliation{Department of
Mathematics, Aliah University, Sector - V , Salt Lake,  Kolkata
700091, India}

 \author{Indrani Karar}
\email{indrani.karar08@gmail.com} \affiliation{Department of Basic
Science and Humanities, Saroj Mohan Institute of Technology,
Guptipara, West Bengal, India.}

\author{Jayanta Naskar}
\email{jayanta86@gmail.com} \affiliation{Department of Physics,
Purulia Zilla School, Purulia, West Bengal, India}

\date{\today}

\begin{abstract}
Stars can be treated as self-gravitating fluid. Krori and Barua
\cite{Krori1975} gave an analytical solutions to that kind of
fluids. In this connection, we propose a de-Sitter model for an
anisotropic strange star with the Krori-Barua spacetime. We
incorporate the existence of cosmological constant in a small
scale to study the structure of anisotropic strange stars and come
to conclusion that this doping is very much compatible with the
well known physical features of strange stars.
\end{abstract}

\maketitle

\section{Introduction}
Recent observational data and results in modern cosmology revealed
that the dark energy which is described in majority by the
cosmological constant $\Lambda$ is of dominant importance in the
dynamics of our Universe. Measurements conducted by Wilkinson
Microwave Anisotropic Probe (WMAP) indicate that almost three
fourth of total mass-energy in the Universe is Dark Energy
\cite{Perlmutter1998,Riess2004} and the leading theory of dark
energy is based on the cosmological constant characterized by
repulsive pressure which was introduced by Einstein in 1917 to
obtain a static cosmological model. Later on Zel'dovich
\cite{Zel'dovich1967} interpreted this quantity physically as a
vacuum energy of quantum fluctuation whose size is of the order of
$\sim 3 \times 10^{-56}$~cm$^{-2}$
\cite{Peebles2003,Padmanabhan2003}.

However, for viability of the present-day accelerated Universe
with dark energy the erstwhile cosmological constant $\Lambda$, in
general, assumed to be time-dependent in the cosmological realm
\cite{Perlmutter1998,Riess2004}. On the other hand,
space-dependent $\Lambda$ has an expected effect in the
astrophysical context as argued by several authors \cite{Chen1990,
Narlikar1991,Ray1993,Tiwari1996} in relation to the nature of
local massive objects like galaxies and elsewhere. In the present
context of compact stars, however, we assume the dark energy in
the form of Einstein's cosmological constant as a purely constant
quantity as follows: either $\Lambda_{eff} = \Lambda_0 -
\Lambda(r)$, where $\Lambda_{eff}$ is the effective cosmological
parameter \cite{Ray2004} or $\Lambda_{eff} = \Lambda_0 + 8\pi E$,
where $E$ is the energy density of the energy state
\cite{Demir2009} so that, for the time being, variation of time
and/or space-dependence of $\Lambda$ is ignored. This constancy of
$\Lambda$ can not be ruled out for the systems of very small
dimension like compact star systems or elsewhere with different
physical requirements
\cite{Mak2000,Dymnikova2002,Dymnikova2003,Bohmer2005}.

To study mass and radii of neutron star Egeland \cite{Egeland2007}
incorporated the existence of cosmological constant
proportionality depending on the density of vacuum. Egeland did it
by using the Fermi equation of state together with the
Tolman-Oppenheimer-Volkov (TOV) equation. Motivated by the above
facts we incorporate the existence of cosmological constant in a
small scale to study the structure of strange stars and arrived to
a conclusion that incorporation of $\Lambda$ describes the well
known strange stars for examples strange stars -  X ray buster,
$4U 1820-30$, X ray pulsar Her $X-1$, Milllisecond pulsar SAX J
$1808.4-3658$ etc., in good manners. Dey et al. \cite{Dey1998},
Usov \cite{Usov2004}, Ruderman \cite{Ruderman1972}, Mak and Harko
\cite{Mak2002,Mak2003,Mak2004}, Li et al. \cite{Li1999a,Li1999b},
Chodos et al. \cite{Chodos1974} and many more have also studied
structure of strange stars in different way. If we look at the
anisotropy and TOV equation of strange stars then our model fit
appropriately with the above said stars.

In the present work, we modelled a strange star which have been
proposed by Alcock et al. and Haensel et al. \cite{Alcock,Haensel}
through their fundamental works. However, our model of strange
star is associated with cosmological constant which satisfies all
the energy conditions including the TOV-equation. We have  checked
the stability and  mass-radius relation. Finally, we have
calculated  the surface redshift  for our solutions which may be
interesting to the observers for possible detection of strange
stars.

\section{ANISOTROPIC de-SITTER MODEL}
To describe the space-time of the strange stars stellar
configuration, we take the Krori and Barua \cite{Krori1975} metric
(henceforth KB) given by
\begin{equation}
ds^2=-e^{\nu(r)}dt^2 + e^{\lambda(r)}dr^2 +r^2
(d\theta^2 +sin^2\theta d\phi^2), \label{eq1}
\end{equation}
with $\lambda(r)=Ar^2$ and $\nu(r) = Br^2 + C$ where $A$, $B$ and
$C$ are arbitrary constants to be determined on  physical grounds.
We further assume that the energy-momentum tensor for the strange
matter filling the interior of the star may be expressed in the
standard form as
\[T_{ij}=diag(\rho,-p_r,-p_t,-p_t)\]
where $\rho$, $p_r$ and $p_t$ correspond to the energy density,
radial pressure and transverse pressure of the baryonic matter,
respectively.

The Einstein's field equations for the metric (\ref{eq1}) in
presence of $\Lambda$ are then obtained as (with $G = c=1$
under geometrized relativistic units)
\begin{eqnarray}
\label{eq2}
 8\pi\rho+ \Lambda &=&
e^{-\lambda}\left(\frac{\lambda^\prime}{r}-\frac{1}{r^2}\right) +
\frac{1}{r^2},\\
\label{eq3}
8\pi p_r - \Lambda &=&
e^{-\lambda}\left(\frac{\nu^\prime}{r}+\frac{1}{r^2}\right) -
\frac{1}{r^2},\\
\label{eq4}
8\pi\ p_t - \Lambda &=&
\frac{e^{-\lambda}}{2}\left[\frac{{\nu^\prime}^2 -
\lambda^{\prime}\nu^{\prime}}{2}
+\frac{\nu^\prime-\lambda^\prime}{r}+\nu^{\prime\prime}\right].
\end{eqnarray}

Now, from the metric (\ref{eq1}) and equations (\ref{eq2}) -
(\ref{eq4}), we get the energy density ($\rho$), the radial
pressure ($p_{r}$) and the tangential pressure ($p_{t}$) as
\begin{eqnarray}
\rho &=&
\frac{1}{8\pi}\left[e^{-Ar^2}\left(2A-\frac{1}{r^2}\right)+\frac{1}{r^2}-\Lambda\right],\label{eq5}\\
p_{r} &=& \frac{1}{8\pi}\left[e^{-Ar^2}\left(2B+\frac{1}{r^2}\right)-\frac{1}{r^2}+\Lambda\right],\label{eq6}\\
p_{t} &=& \frac{1}{8\pi}\left[e^{-Ar^2}\left((B^2
-AB)r^2+(2B-A)\right)+\Lambda\right].\label{eq7}
\end{eqnarray}

Using equations (\ref{eq5}) - (\ref{eq7}) the equation of state
(EOS) corresponding to radial and transverse directions may be
written as
\begin{equation}
\label{eq11} \omega_r(r)   =
\frac{\left[e^{-Ar^2}\left(2B+\frac{1}{r^2}\right)-\frac{1}{r^2}+\Lambda\right]
}{\left[e^{-Ar^2}\left(2A-\frac{1}{r^2}\right)+\frac{1}{r^2}-\Lambda\right]},
\end{equation}
\begin{equation}
\label{eq12} \omega_t(r)   =
\frac{e^{-Ar^2}\left[\left(B^2-AB\right)r^2+(2B-A)\right]+\Lambda
}{\left[e^{-Ar^2}\left(2A-\frac{1}{r^2}\right)+\frac{1}{r^2}-\Lambda\right]}.
\end{equation}

\section{PHYSICAL ANALYSIS}
It is known that $\Lambda > 0$ implies the space is open. To
explain the present acceleration state of the universe, it is
believed that energy in the vacuum is responsible for this
expansion. As a consequence, Vacuum energy provides some
gravitational effect on the stellar structures. It is suggested
that cosmological constant plays the role of  energy of the
vacuum. In this section we will study the following features of
our model assuming the value of $\Lambda = 0.00018$ km$^{-2}$. We have
assumed this value  as required for the  stability of the strange
star and mathematical consistency.

\subsection{Anisotropic Behavior} From the equation (\ref{eq5})
we have,
\[ \frac{d\rho}{dr}= -\frac{1}{8\pi}\left[\left(4A^2 r
- \frac{2A}{r} -\frac{2}{r^3}\right)e^{-Ar^2}+\frac{2}{r^3}\right] < 0,\]\\
and equation (6) leads to
\[\frac{dp_r}{dr} < 0.\]

The density and pressure are decreasing with the increase of
radius of the star.

  Figs. 1 and 2 support the above results.

\begin{figure}[htbp]
    \centering
        \includegraphics[scale=.3]{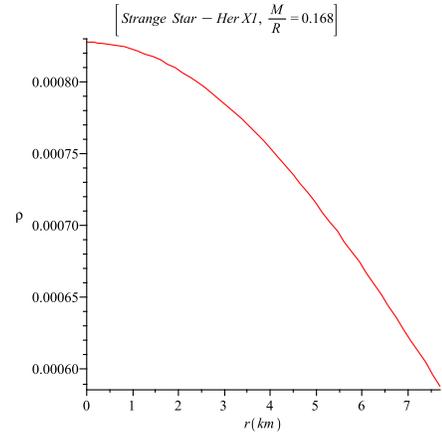}
        \caption{ Density variation at the stellar interior of Strange Star-Her X-1}
    \label{fig:1}
\end{figure}

\begin{figure}[htbp]
    \centering
        \includegraphics[scale=.3]{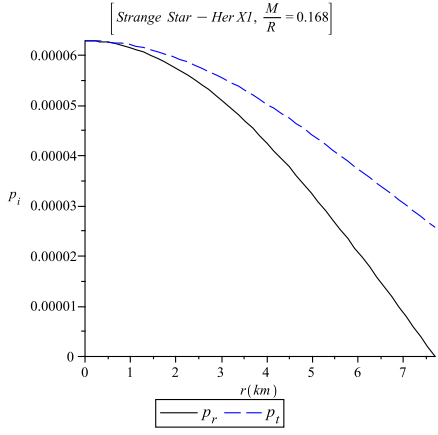}
        \caption{ Radial and Transverse Pressure  variation at the stellar interior of Strange Star-Her X-1}
    \label{fig:2}
\end{figure}

We observe that,
 at $r=0$, our model provides\\
\[ \frac{d\rho}{dr}=0,~~ \frac{dp_r}{dr} = 0, \]
\[ \frac{d^2 \rho}{dr^2}=-\frac{A^2}{4\pi} < 0,\]
and
\[\frac{d^2 p_r}{dr^2} < 0,\]
which indicate maximality of central density and central pressure.
Interestingly, similar to an ordinary matter distribution, the EOS
is restricted between 0 and 1  (see Fig. 3) despite the fact that
star is constituted by the combination of strange matter and
effect of $\Lambda$. We call it as strange matter as EOS depends
on radius of the star rather a constant as in ordinary matter.

\begin{figure}[htbp]
    \centering
        \includegraphics[scale=.3]{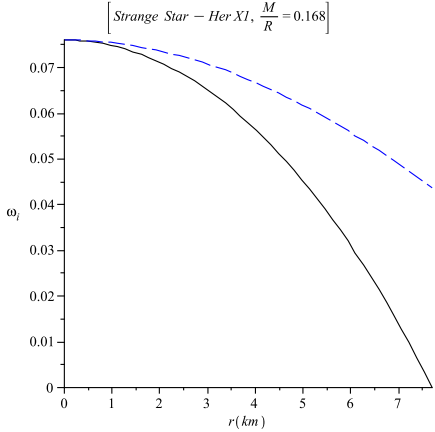}
        \caption{ Variation of equation of state parameter with distance of Strange Star-Her X-1}
    \label{fig:3}
\end{figure}

The measure of anisotropy, $\Delta  =
\left(p_{t}-p_{r}\right)$ in this model is obtained as
\begin{multline}
\label{eq13} \Delta  =
 \frac{1}{8\pi
}\left[e^{-Ar^2}\left((B^2 -AB)r^2
-A-\frac{1}{r^2}\right)+\frac{1}{r^2}\right].
\end{multline}

We note that measure of anisotropy is independent of $\Lambda$. In
other words, vacuum energy does not affect on the anisotropic
force. The `anisotropy' will be directed outward when $P_t > P_r$
i.e. $\Delta > 0 $, and inward if $P_t < P_r$ i.e. $\Delta < 0$.
  Fig. 4 of our model indicates that $\Delta > 0$ i.e.  a repulsive
`anisotropic' force exists for Strange Star-4U 1820-30, Her X-1
and  SAX J 1808.4-3658.  The positivity of $\Delta$ allows the
construction of more massive distributions.

\begin{figure}[htbp]
\centering
\includegraphics[scale=.3]{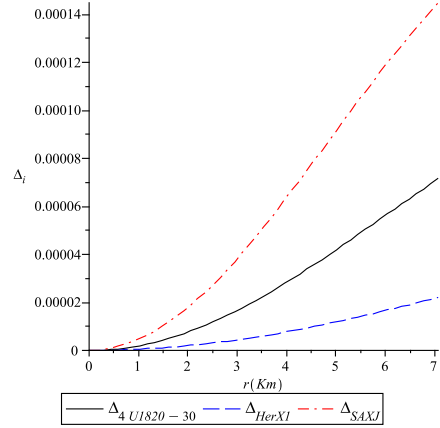}
\caption{Comparison of anisotropic behaviours  at the stellar
interior of Strange Star-4U 1820-30, Her X-1 and  SAX J
1808.4-3658.}
    \label{fig:4}
\end{figure}

\subsection{Matching Conditions}
Here we match the interior metric to the Schwarzschild de Sitter
exterior
\begin{multline}
ds^{2}=-\left(1 - \frac{2M}{r}- \frac{\Lambda r^2}{3}  \right)dt^2
+ \left(1 - \frac{2M}{r}- \frac{\Lambda r^2}{3}
\right)^{-1}dr^2 \\+ r^2(d\theta^2+\sin^2\theta d\phi^2),
\label{eq20}
\end{multline}
at the boundary $r= R$. Continuity of the metric functions
$g_{tt}$, $g_{rr}$ and $\frac{\partial g_{tt}}{\partial r}$ at the
boundary surface $S$ yields

\begin{eqnarray}
A &=& - \frac{1}{R^2} \ln \left[ 1 - \frac{2M}{R}-\frac{1}{3}\Lambda R^2 \right],\label{eq15}\\
B &=& \frac{1}{R} \left[\frac{M}{R^2}-\frac{1}{3}\Lambda R\right]
\left[1 -\frac{2M}{R} -\frac{1}{3}\Lambda R^2  \right]^{-1}, \label{eq16}\\
C &=&  \ln \left[ 1 - \frac{2M}{R}-\frac{1}{3}\Lambda
R^2\right]-\frac{ R\left[\frac{M}{R^2}-\frac{1}{3}\Lambda R\right]
 }{ \left[ 1 - \frac{2M}{R}-\frac{1}{3}\Lambda R^2
\right]}. \label{eq17}
\end{eqnarray}

\begin{table*}
\centering
\begin{minipage}{140mm}
\caption{Values of the model parameters for
different Strange stars.}\label{tbl-1}
\begin{tabular}{@{}lrrrrrr@{}}
\hline
Strange Quark Star & $M$($M_{\odot}$) & $R$(km) & $\frac{M}{R}$ & $A$ (km$^{-2}$) & $B$ (km$^{-2}$) & b (km$^{-2}$) \\ \hline
Her X-1 & 0.88 & 7.7 & 0.168 & 0.0069968808 & 0.004199502132 & 0.000828449  \\
SAX J 1808.4-3658(SS1) & 1.435 & 7.07 & 0.299 & 0.0138138300 & 0.01484158661 & 0.002188063 \\
4U 1820-30 & 2.25 & 10.0 & 0.332 & 0.01108662625 & 0.009878787879 & 0.001316874\\
\hline
\end{tabular}
\end{minipage}
\end{table*}

Imposing the boundary conditions $p_r(r=R) = 0$ and $\rho( r=0) =
b$ (=a constant), where $b$ is the central density, we have A and
B in the following  forms:
\begin{eqnarray}
A &=& \frac{8 \pi b + \Lambda}{3},\label{eq18} \\
B &=& \frac{1}{2} \left[ e^{\frac{8 \pi b + \Lambda}{3}
R^2}\left(\frac{1}{R^2}-\Lambda\right) -\frac{1}{R^2}
\right].\label{eq19}
\end{eqnarray}

Combining, equations (\ref{eq15}) and (\ref{eq18}), we get
\begin{equation}
A = \frac{8 \pi b+ \Lambda}{3} = - \frac{1}{R^2} \ln \left[ 1 -
\frac{2M}{R}-\frac{1}{3}\Lambda R^2 \right]. \label{eq20}
\end{equation}

At this juncture, to get an insight of our model, we have
evaluated the numerical values of the parameters A, B and b for
the Strange Star-4U 1820-30, Her X-1 and  SAX J 1808.4-3658 ( see
table 1 ).

We have verified for particular choices of the values of mass and
radius leading to solutions for the unknown parameters, satisfy
the following energy conditions namely, the null energy condition
(NEC), weak energy condition (WEC), strong energy condition (SEC)
and dominant energy condition (DEC) through out the configuration:
\[\rho \geq 0, \rho+p_r \geq 0,\rho+p_t \geq 0,\rho+p_r+2p_t \geq 0,\rho > |p_r|,\]
and $\rho > |p_t|.$\\
\\

It is interesting to note here that the model satisfies the strong
energy condition, which implies that the space-time does contain a
black hole region.

The anisotropy, as expected, vanishes at the centre i.e., $p_t =
p_r = p_0=\frac{2B-A+\Lambda}{8\pi}$ at r=0. The  energy density
and the two pressures are also well behaved in the interior of the
stellar configuration.

\subsection{TOV Equation}
For an anisotropic fluid distribution, the generalized TOV
equation is given by
\begin{equation}
\label{eq24} \frac{d}{dr}\left(p_r -\frac{\Lambda}{8\pi} \right)
+\frac{1}{2} \nu^\prime\left(\rho +p_r\right) +
\frac{2}{r}\left(p_r - p_t\right) = 0.
\end{equation}
According to Ponce de Le\'{o}n \cite{Leon1993}, the above TOV
equation can be rewritten as
\begin{multline}
-\frac{M_G\left(\rho+p_r\right)}{r^2}e^{\frac{\lambda-\nu}{2}}-\frac{d}{dr}
\left(p_r -\frac{\Lambda}{8\pi} \right)\\
+\frac{2}{r}\left(p_t-p_r\right)=0, \label{eq25}
\end{multline}
where $M_G=M_G(r)$ is the gravitational mass inside a
sphere of radius $r$ and is given by
\begin{equation}
M_G(r)=\frac{1}{2}r^2e^{\frac{\nu-\lambda}{2}}\nu^{\prime},\label{eq26}
\end{equation}
which can easily be derived from the Tolman-Whittaker formula and
the Einstein's field equations. This new form of TOV equation
provides the equilibrium condition for the strange star  subject
to gravitational and hydrostatic plus another force due to the
anisotropic nature of the stellar object. Using equations
(\ref{eq5}) - (\ref{eq7}), the above equation can be written as
\begin{equation}
 F_g+ F_h + F_a=0,\label{eq27}
\end{equation}
where,
\begin{eqnarray}
F_g &=& -B r\left(\rho+p_r\right),\label{eq28}\\
F_h &=& -\frac{d}{dr}\left(p_r -\frac{\Lambda}{8\pi} \right),\label{eq29}\\
F_a &=& \frac{2}{r}\left(p_t -p_r\right).\label{eq30}
\end{eqnarray}
The profiles of $F_g$, $F_h$ and $F_a$ for our chosen source are
shown in Fig. 5. This figure indicates that the static equilibrium
can be attained due to pressure anisotropy, gravitational and
hydrostatic forces.

\begin{figure}[htbp]
\centering
\includegraphics[scale=.3]{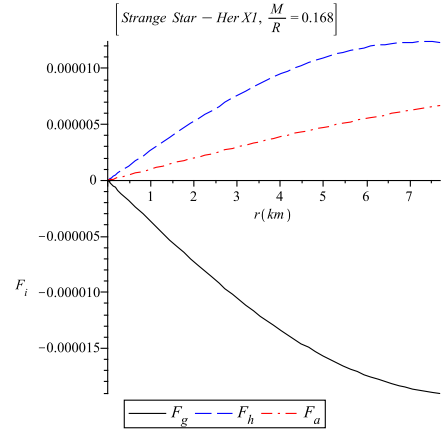}
\caption{Behaviours of pressure anisotropy,gravitational and
hydrostatic forces at the stellar interior of Strange Star  Her
X-1 .}
    \label{fig:5}
\end{figure}

\subsection{Stability}
The velocity of sound $ v_s^2=(\frac{dp}{d\rho}) $ should be less
than one for a realistic model. \cite{Herrera1992,Abreu2007}. Now,
we calculate the radial and transverse speed for  our anisotropic
model,
\begin{multline}
\label{eq31}
v_{sr}^{2}=\frac{dp_r}{d\rho}\\=-1+\frac{4Are^{-Ar^2}(A+B)}{e^{-Ar^2}
\left(4A^2r-\frac{2A}{r}-\frac{2}{r^3}\right)+\frac{2}{r^3}},
\end{multline}
\begin{multline}
\label{eq32}
v^2_{st}=\frac{dp_t}{d\rho}\\=-\frac{e^{-Ar^2}\left[2r\left(A^2
+B^2-3AB\right)-2A(B^2-AB)r^3\right]}{e^{-Ar^2}\left(4A^2r-\frac{2A}{r}
-\frac{2}{r^3}\right)+\frac{2}{r^3}}
\end{multline}
To check whether the sound speeds lie between 0 and 1, we plot the
radial and transverse sound speeds in Fig. 6 and observe that
these parameters satisfy the inequalities $0\leq v_{sr}^2 \leq 1$
and $0\leq v_{st}^2 \leq 1$ everywhere within the stellar object.

Equations (\ref{eq31}) and (\ref{eq32}) lead to
\begin{multline}
\label{eq33}
v^2_{st}-v^2_{sr}\\=1-\frac{e^{-Ar^2}\left[2r\left(3A^2
+B^2-AB\right)+2AB(A-B)r^3\right]}{e^{-Ar^2}\left(4A^2r-\frac{2A}{r}-\frac{2}{r^3}\right)+\frac{2}{r^3}}
\end{multline}
As sound speeds lie between 0 and 1, therefore,  $\mid v_{st}^2 -
v_{sr}^2 \mid \leq 1 $.

In few years back, Herrera's \cite{Herrera1992} proposed a
technique for stability check of local anisotropic matter
distribution. This technique is known as cracking (or overturning)
concept which states that the region for which radial speed of
sound is greater than the transverse speed of sound  is a
potentially stable region.
 In our case, Fig.~7 indicates that
there is no change of sign for the term $v_{st}^2 - v_{sr}^2 $
within the specific configuration. Also, the plot for $v_{st}^2 -v_{sr}^2$
(Fig. 7) shows negativity in its nature. Therefore, we conclude
that our strange star model is stable.

\begin{figure}[htbp]
\centering
\includegraphics[scale=.3]{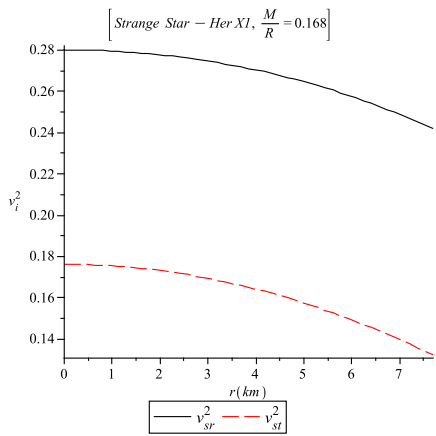}
\caption{ Variation of radial and transverse sound speed of
Strange Star-Her X-1}
    \label{fig:6}
\end{figure}

\begin{figure}[htbp]
\centering
\includegraphics[scale=.3]{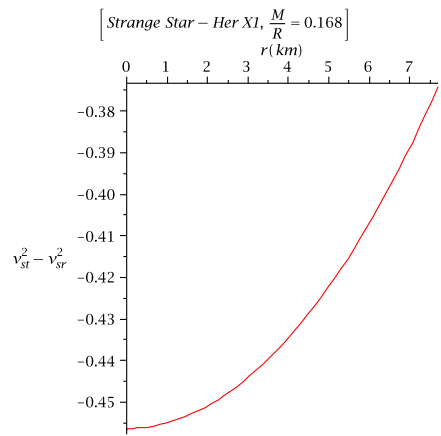}
\caption{ Variation of the $v_{st}^2-v_{sr}^2$ of Strange Star-Her
X-1}
    \label{fig:7}
\end{figure}

\subsection{Surface redshift}
To get observational evidence of anisotropies in the internal
pressure distribution, it is necessary to study   redshift of
light emitted at the surface of the compact objects. At first, we
try to see whether our model will follow Buchdahl
\cite{Buchdahl1959} maximum allowable mass radius ratio limit. We
have calculated $\frac{M_{eff}} {R} = \frac{4 \pi \int_0^R \rho dr
} {R}$ for Strange Star  Her X-1 of our model  and have found that
$\left(\frac{M_{eff}} {R}\right)_{max} = .336$. Thus our model
satisfies Buchdahl's limit and hence physically acceptable. It is
worthwhile to mention that our model provides the same mass radius
ratio for the observed Strange Star  Her X-1.

 The compactness of the star is
given by
\begin{equation}
\label{eq35} u= \frac{M_{eff}} {R}= \frac{1}{2}\left( 1-e^{-AR^2}
\right)- \frac{\Lambda R^2}{6}.
\end{equation}
The surface redshift ($Z_s$) corresponding to the above
compactness ($u$) is obtained as
\begin{equation}
\label{eq36} 1+Z_s= \left[ 1-(2 u+ \frac{\Lambda
R^2}{3})\right]^{-\frac{1}{2}} ,
\end{equation}
where
\begin{equation}
\label{eq37} Z_s=  e^{ \frac{A}{2}R^2}-1.
\end{equation}
Thus, the maximum surface redshift for a strange star Her X-1 of
radius $7.7~$km turns out to be $Z_s = 0.022$  (see Fig. 8).

\begin{figure}[htbp]
\centering
\includegraphics[scale=.3]{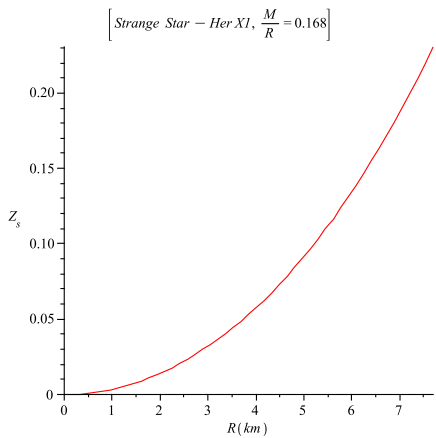}
\caption{Variation of surface redshift of Strange Star-Her X-1}
\label{fig:8}
\end{figure}

\section{Discussion}
We have studied in the present work a self-gravitating fluid of
strange star under the metric of KB \cite{Krori1975}. The spacetime turns
out to be de Sitter type and anisotropic in nature due to the
presence of tangential pressure. We have incorporated the
erstwhile cosmological constant $\Lambda$ in the field equation to
study the structure of anisotropic strange stars. Successfully we
find out an analytical solution to the fluids which are quite
interesting in connection to several physical features of strange
stars.

In this regard it is to note that the present investigation is a
sequel of the earlier works with KB models
\cite{Varela2010,Rahaman2010,Rahaman2011,Rahaman2011a}. Varela et
al. \cite{Varela2010} considered KB model in the context of
electrically charged system whereas Rahaman et al.
\cite{Rahaman2010} studied it under the influence of anisotropic
charged fluids with Chaplygin equation of state. Also, Rahaman
et al. \cite{Rahaman2011} studied dark energy star constituted by
two fluids, namely ordinary fluid and dark energy. Very recently
Rahaman et al. \cite{Rahaman2011a} have uniquely considered the
system of strange star with MIT Bag model under KB metric.
  However, in the present investigation we have considered
a de-Sitter model for an anisotropic strange star with the same KB
\cite{Krori1975} spacetime. It is yet unclear to what extent the
strange stars can be described by the KB \cite{Krori1975} metric.
At least, at this stage of theoretical investigation on strange
star, the question of `if they exist at all' is not a serious
issue or adding one more arbitrary ingredient to match with the
observational data can not seriously disprove the subject.

We would like to mention here that we have doped here $\Lambda$ as
a purely constant quantity and have shown that the results are
very much compatible with the well known physical features of
strange stars. This at once demands a space-variable $\Lambda$
\cite{Chen1990, Narlikar1991,Ray1993,Tiwari1996} to be
incorporated to see the effect of this inclusion in the
astrophysical system like strange star.   Another issue, possibly
very intriguing one, the assumed value of the cosmological
constant here as $\Lambda = 0.00018$ km$^{-2}$. This has certainly
nothing to do with the standard ``cosmology'' of mainstream arena
as it looks quite artificial. However, all these issues related to
$\Lambda$ may be considered in a future project.

\section*{Acknowledgments}
MK, FR and SR gratefully acknowledge support from IUCAA, Pune,
India for providing Visiting Associateship under which a part of
this work was carried out. SMH thanks to IUCAA for giving him an
opportunity to visit IUCAA where a part of this work was carried
out. FR is also thankful to UGC for providing financial support.
Finally, we are expressing our deep gratitude to the anonymous
referee for  suggesting some pertinent issues that have led to
significant improvements.


\begin{thebibliography}{99}

\bibitem[\protect\citeauthoryear{Krori}{1975}]{Krori1975}
K.D. Krori and J. Barua, J. Phys. A.: Math. Gen. {\bf 8}, 508
(1975).

\bibitem[\protect\citeauthoryear{Perlmutter}{1998}]{Perlmutter1998}
 S. Perlmutter et al., Nature {\bf 391}, 51 (1998).

\bibitem[\protect\citeauthoryear{Riess}{2004}]{Riess2004} A.G. Riess et al., Astrophys. J. {\bf 607}, 665 (2004).

\bibitem[\protect\citeauthoryear{Zel'dovich}{1967}]{Zel'dovich1967} Y. B. Zel'dovich, JETP letters {\bf 6}, 316 (1967);
Sov. Phys. Uspekhi {\bf 11}, 381 (1968).

\bibitem[\protect\citeauthoryear{Peebles}{2003}]{Peebles2003} P.J.E. Peebles and B. Ratra, Rev. Mod. Phys. {\bf 75}, 559 (2003).

\bibitem[\protect\citeauthoryear{Padmanabhan}{2003}]{Padmanabhan2003} T. Padmanabhan, Phys. Rep. {\bf 380}, 235 (2003).

\bibitem[\protect\citeauthoryear{Chen}{1990}]{Chen1990} W. Chen and Y.S. Wu, Phys. Rev. D {\bf 41}, 695 (1990).

\bibitem[\protect\citeauthoryear{Narlikar}{1991}]{Narlikar1991} J.V. Narlikar, J.-C Pecker, J.-P Vigier, J.
Astrophys. Astr. {\bf 12}, 7 (1991).

\bibitem[\protect\citeauthoryear{Ray}{1993}]{Ray1993} S. Ray and D. Ray, Astrophys. Space Sci. {\bf 203}, 211 (1993).

\bibitem[\protect\citeauthoryear{Tiwari}{1996}]{Tiwari1996} R.N. Tiwari and S. Ray, Ind. J. Pure Appl. Math. {\bf 27}, 907 (1996).

\bibitem[\protect\citeauthoryear{Ray}{2004}]{Ray2004} S. Ray and S. Bhadra, Phys. Lett. A {\bf 322}, 150
(2004).

\bibitem[\protect\citeauthoryear{Demir}{2009}]{Demir2009} D.A. Demir, Found. Phys. {\bf 39}, 1407
(2009).

\bibitem[\protect\citeauthoryear{Mak}{2000}]{Mak2000} M. K. Mak and P.N. Dobson, JR., Mod. Phys. Lett. A {\bf
15}, 2153 (2000).

\bibitem[\protect\citeauthoryear{Dymnikova}{2002}]{Dymnikova2002} I. Dymnikova, Class. Quantum Grav. {\bf 19} 725 (2002).

\bibitem[\protect\citeauthoryear{Dymnikova}{2003}]{Dymnikova2003} I. Dymnikova, Int. J. Mod. Phys. D {\bf
12}, 1015 (2003).

\bibitem[\protect\citeauthoryear{Bohmer}{2005}]{Bohmer2005} C.G. B{\"o}hmer and T. Harko, Physics Letters B 630 (2005) 73

\bibitem[\protect\citeauthoryear{Egeland}{2007}]{Egeland2007} E. Egeland, {\it Compact Stars} (Trondheim, Norway, 2007).

\bibitem[\protect\citeauthoryear{Dey}{1998}]{Dey1998} M. Dey et al., Phys. Lett. B. {\bf 438}, 123 (1998),
addendum:{\bf 447}, 352 (1999), erratum;{\bf 467}, 303 (1999).

\bibitem[\protect\citeauthoryear{Usov}{2004}]{Usov2004} V.V.
Usov, Phys. Rev. D. {\bf 70}, 067301 (2004).

\bibitem[\protect\citeauthoryear{Ruderman}{1972}]{Ruderman1972} R.
Ruderman, Ann. Rev. Astron. Astrophys. {\bf 10}, 427 (1972).

\bibitem[\protect\citeauthoryear{Mak}{2002}]{Mak2002}
M.K. MaK and T. Harko, Chin. J. Astron. Astrophys. {\bf 2}, 248
(2002).

\bibitem[\protect\citeauthoryear{Mak}{2003}]{Mak2003}
M.K. Mak and T. Harko, Proc. R. Soc. Lond. {\bf 459}, 393 (2003).

\bibitem[\protect\citeauthoryear{Mak}{2004}]{Mak2004}
M.K. Mak and T. Harko, Int. J. Mod. Phys. D. {\bf 13}, 149 (2004).

\bibitem[\protect\citeauthoryear{Li}{1999a}]{Li1999a}
X.-D. Li et al., Phys. Rev. Lett. {\bf 82}, 3776 (1999).

\bibitem[\protect\citeauthoryear{Li}{1999b}]{Li1999b}
X.-D. Li et al., Astrophys. J. {\bf 527}, L51 (1999).

\bibitem[\protect\citeauthoryear{Chodos}{1974}]{Chodos1974}
A. Chodos et al., Phys. Rev. D. {\bf 9}, 3471 (1974).

\bibitem[\protect\citeauthoryear{Alcock et al.}{1986}]{Alcock}
C. Alcock, E. Farhi and A. Olinto, Astrophys. J. {\bf 310}, 261 (1986).

\bibitem[\protect\citeauthoryear{Haensel et al.}{1986}]{Haensel}
P. Haensel, J. L. Zdunik and R. Schaeffer, Astron. Astrophys. {\bf 160}, 121 (1986).

\bibitem[\protect\citeauthoryear{Buchdahl}{1959}]{Buchdahl1959}
H.A. Buchdahl, Phys. Rev. {\bf 116}, 1027 (1959).

\bibitem[\protect\citeauthoryear{leon}{1993}]{Leon1993}
J. Ponce de Le\'{o}n, Gen. Relativ. Gravit. {\bf 25}, 1123 (1993).

\bibitem[\protect\citeauthoryear{Herrera}{1992}]{Herrera1992}
L. Herrera, Phys. Lett. A, {\bf 165} 206, (1992).

\bibitem[\protect\citeauthoryear{Abreu}{2007}]{Abreu2007} H. Abreu,
H. Hernandez and L.A. Nunez, Class. Quantum Gravit. {\bf 24}, 4631
(2007).

\bibitem[\protect\citeauthoryear{Varela et al.}{2010}]{Varela2010}
V. Varela, F. Rahaman, S. Ray, K. Chakraborty and M. Kalam, Phys. Rev. D {\bf 82}, 044052 (2010).

\bibitem[\protect\citeauthoryear{Rahaman et al.}{2010}]{Rahaman2010}
F. Rahaman, S. Ray, A. K. Jafry and K. Chakraborty, Phys. Rev. D {\bf 82} 104055 (2010).
\bibitem[\protect\citeauthoryear{Rahaman et al.}{2011}]{Rahaman2011}
F. Rahaman,  R. Maulick , A. K. Yadav,  S. Ray, R. Sharma,
Gen.Rel.Grav. {\bf 44}  107 (2012)
 arXiv:1102.1382 [gr-qc]
\bibitem[\protect\citeauthoryear{Rahaman et al.}{2011}]{Rahaman2011a}
F. Rahaman, R. Sharma, S. Ray, R. Maulick and I. Karar,
arxiv:1108.6125[gr-qc] (2011) ( to appear in Eur.Phys.J. C
(2012)).

\end{thebibliography}
\end{document}